\documentclass[onecolumn]{aastex631}

\usepackage{xcolor}
\usepackage{CJKutf8}

\usepackage{natbib} 
\usepackage{amsmath}
\usepackage{enumitem}
\shorttitle{DEEP III - Survey Simulation}
\shortauthors{Bernardinelli et al.}

\newcommand{\E}{\mathrm{E}}

\newcommand\change[1]{#1}

\def \NAU {Department of Astronomy and Planetary Science, Northern Arizona University,\\PO Box 6010, Flagstaff, AZ 86011, USA}
\def \UMPhysics {Department of Physics, University of Michigan,\\ Ann Arbor, MI 48109, USA}
\def \UMAstronomy {Department of Astronomy, University of Michigan,\\ Ann Arbor, MI 48109, USA}
\def \UW {DiRAC Institute and the Department of Astronomy, University of Washington, Seattle, USA}
\def \uchile {Departamento de Astronomía, Universidad de Chile,\\ Camino del Observatorio 1515, Las Condes, Santiago, Chile}
\def \cfa {Harvard-Smithsonian Center for Astrophysics,\\ 60 Garden St., MS 51, Cambridge, MA 02138, USA}
\def \byu {Department of Physics and Astronomy, Brigham Young University, Provo, UT 84602, USA}
\def \apl {Applied Physics Lab, Johns Hopkins University,\\ 11100 Johns Hopkins Road, Laurel, Maryland 20723, USA}
\def \ucla {Department of Earth, Planetary and Space Sciences, University of California Los Angeles, 595 Charles E. Young Dr. East, Los Angeles, CA 90095, USA}
\def \carnegie {Earth and Planets Laboratory, Carnegie Institution for Science, Washington, DC 20015}
\def \stgallen {School of Computer Science, University of St. Gallen,\\ Rosenbergstrasse 30, CH-9000 St. Gallen, Switzerland}

\begin{document} 

\title{The DECam Ecliptic Exploration Project (DEEP) III: Survey characterization and simulation methods}

\author[0000-0003-0743-9422]{Pedro H. Bernardinelli}
\altaffiliation{DiRAC Postdoctoral Fellow}
\affiliation{\UW}
\email{phbern@uw.edu}
\author{Hayden Smotherman} 
\affiliation{\UW}

\author[0000-0001-7574-4440]{Zachary Langford}
\affiliation{Department of Physics and Astronomy, University of Pennsylvania, Philadelphia, PA, USA}
\affiliation{\UW}

\author[0000-0001-8132-8056]{Stephen K. N. Portillo}
\affiliation{Department of Mathematical and Physical Sciences, Concordia University of Edmonton, 7128 Ada Boulevard, Edmonton, AB, T5B 4E4 Canada}
\affiliation{\UW}

\author{Andrew J. Connolly} 
\affiliation{\UW}
\author{J. Bryce Kalmbach}
 \affiliation{\UW}
\author{Steven Stetzler} 
\affiliation{\UW}

\author[0000-0003-1996-9252]{Mario Juri\'c}
\affiliation{\UW}

\author[0000-0001-5750-4953]{William J. Oldroyd}
\affiliation{\NAU}
\author[0000-0001-7737-6784]{Hsing~Wen~Lin (\begin{CJK*}{UTF8}{gbsn} 林省文\end{CJK*})}
\affiliation{\UMPhysics}

\author[0000-0002-8167-1767]{Fred C.~Adams}
\affiliation{\UMPhysics}
\affiliation{\UMAstronomy}

\author[0000-0001-7335-1715]{Colin Orion Chandler}
\affiliation{\UW}
\affiliation{LSST Interdisciplinary Network for Collaboration and Computing, 933 N. Cherry Avenue, Tucson AZ 85721}
\affiliation{\NAU}

\author[0000-0002-5211-0020]{Cesar Fuentes}
\affiliation{\uchile}

\author[0000-0001-6942-2736]{David~W.~Gerdes}
\affiliation{\UMPhysics}
\affiliation{\UMAstronomy}

\author[0000-0001-8550-6788]{Matthew J. Holman}
\affiliation{\cfa}

\author[0000-0002-2486-1118]{Larissa Markwardt}
\affiliation{\UMPhysics}

\author{Andrew McNeill}
\affiliation{\NAU}
\affiliation{Department of Physics, Lehigh University, 16 Memorial Drive East, Bethlehem, PA, 18015, USA}

\author[0000-0002-7817-3388]{Michael Mommert}
\affiliation{\stgallen}

\author[0000-0003-4827-5049]{Kevin J. Napier}
\affiliation{\UMPhysics}

\author[0000-0001-5133-6303]{Matthew J. Payne}
\affiliation{\cfa}

\author[0000-0003-1080-9770]{Darin Ragozzine}
\affiliation{\byu}

\author[0000-0002-9939-9976]{Andrew S. Rivkin}
\affiliation{\apl}

\author{Hilke Schlichting}
\affiliation{\ucla}

\author[0000-0003-3145-8682]{Scott S. Sheppard}
\affiliation{\carnegie}

\author[0000-0001-6350-807X]{Ryder Strauss}
\affiliation{\NAU}

\author[0000-0003-4580-3790]{David E. Trilling}
\affiliation{\NAU}

\author[0000-0001-9859-0894]{Chadwick A. Trujillo}
\affiliation{\NAU}

\begin{abstract}
We present a detailed study of the observational biases of the DECam Ecliptic Exploration Project's (DEEP) B1 data release and survey simulation software that enables direct statistical comparisons between models and our data. We inject a synthetic population of objects into the images, and then subsequently recover them in the same processing as our real detections. This enables us to characterize the survey's completeness as a function of apparent magnitudes and on-sky rates of motion. We study the statistically optimal functional form for the magnitude, and develop a methodology that can estimate the magnitude and rate efficiencies for all survey's pointing groups simultaneously. We have determined that our peak completeness is on average 80\% in each pointing group, and our magnitude drops to $25\%$ of this value at $m_{25} = 26.22$. We describe the freely available survey simulation software and its methodology. We conclude by using it to infer that our effective search area for objects at 40 au is $14.8\deg^2$, and that our lack of dynamically cold distant objects means that there at most $8\times 10^3$ objects with $60 < a < 80$ au and absolute magnitudes $H \leq 8$. 
\end{abstract}

\section{Introduction} \label{sec:intro}

Our understanding of the trans-Neptunian region has improved significantly since the discovery of the first trans-Neptunian object (TNO) by \cite{Jewitt1993}. Thousands of objects have now been discovered, individual surveys are now capable of finding hundreds of objects \citep[\emph{e.g.}][]{Petit2011,Bannister2015,Bannister2018,Bernardinelli2019,Bernardinelli2022}, and models of this region make quantitative predictions of the orbital structure, number of objects and size distribution of this population \cite[see reviews by][]{Morbidelli2020,Gladman2021}.  \change{However, such predictions are not immediately testable against the observations from any given survey. Any survey has observational biases, driven primarily by the observational strategy used and the apparent magnitude limits achieved in the exposures. Thus, any comparison of a model to the real trans-Neptunian region (as seen by each survey) must factor in these effects.} 

Survey simulators have become widespread tools for testing these population models against observations \citep{Jones2006,Lawler2018,Bernardinelli2022}. These tools provide a standardized way to estimate discovery biases of a survey for any given synthetic object as a function of orbital elements, magnitude, colors and light curve properties without requiring the need to directly inject objects into the survey images, which can oftentimes be a prohibitively expensive task, in terms of computation time, especially if multiple models are being tested. The detailed understanding of these biases means that one can test not only the populations of detected objects, but also place meaningful limits given survey non-detections in a region of parameter space. Examples in the literature include, but are not limited to, tests for clustering of high \change{semi-major axis ($a$)} and high \change{perihelion ($q$)} trans-Neptunian objects \citep{Shankman2017,Bernardinelli2020,Napier2021}, limits on the discoverability of distant planets in the Solar System \citep{belyakov2022}, tests of the structure of the classical Kuiper belt \citep{Petit2011,Kavelaars2021,Bernardinelli2022} and size distribution calculations \citep{Lawler2018a}.

The DECam Ecliptic Exploration Project \citep[DEEP,][]{DEEPI} is an ongoing survey using the Dark Energy Camera \citep[DECam,][]{Flaugher2015} dedicated to finding faint trans-Neptunian objects with digital tracking, using a carefully chosen observational strategy \citep{DEEPII} that maximizes the discovery potential of low inclination Kuiper belt objects. The survey consists of a series of ``pointing groups'' or ``long stares'', with a nominal observing strategy of 100 images per night in an expanding cone-shaped pattern that follows the typical motion of the TNO population (see Figure 2 of \citet{DEEPII}). The survey consists of four fields (A0, A1, B0 and B1) that started being imaged in 2019. The analysis presented here uses data from the B1 field with data taken between 2019 and 2021. Applications of the techniques presented here to the other DEEP fields are forthcoming. \change{This observational strategy enables the DEEP project to reach a unique combination of depth and area, which will enable us to probe the absolute magnitude distribution of TNOs as a function of sub-population with high statistical significance at magnitudes fainter than $m_{r} \approx 25$, beyond the limits of current \citep{Bannister2018,Bernardinelli2022} and upcoming \citep{ivezic2019} large area surveys \citep[see][for a detailed description of the DEEP survey goals]{DEEPI}.}

In Section \ref{sec:fakes} we present a synthetic population of objects injected into the DEEP images, and in Section \ref{sec:eff} we discuss how this population can be used to characterize the survey's completeness as a function of magnitudes and orbital elements. We then use these results to construct a survey simulator, presented in Section \ref{sec:methods} and publicly released on \texttt{GitHub}\footnote{\url{https://github.com/bernardinelli/DESTNOSIM}}. Section \ref{sec:numbers} presents two quantitative analyses of ``toy models'' of the trans-Neptunian region to demonstrate the power of the survey simulator, and we conclude in Section \ref{sec:conc} by outlining expected improvements.

\section{Synthetic population injected in the DEEP data}
\label{sec:fakes}
\begin{figure}[ht!]
	\centering
	\includegraphics[width=\textwidth]{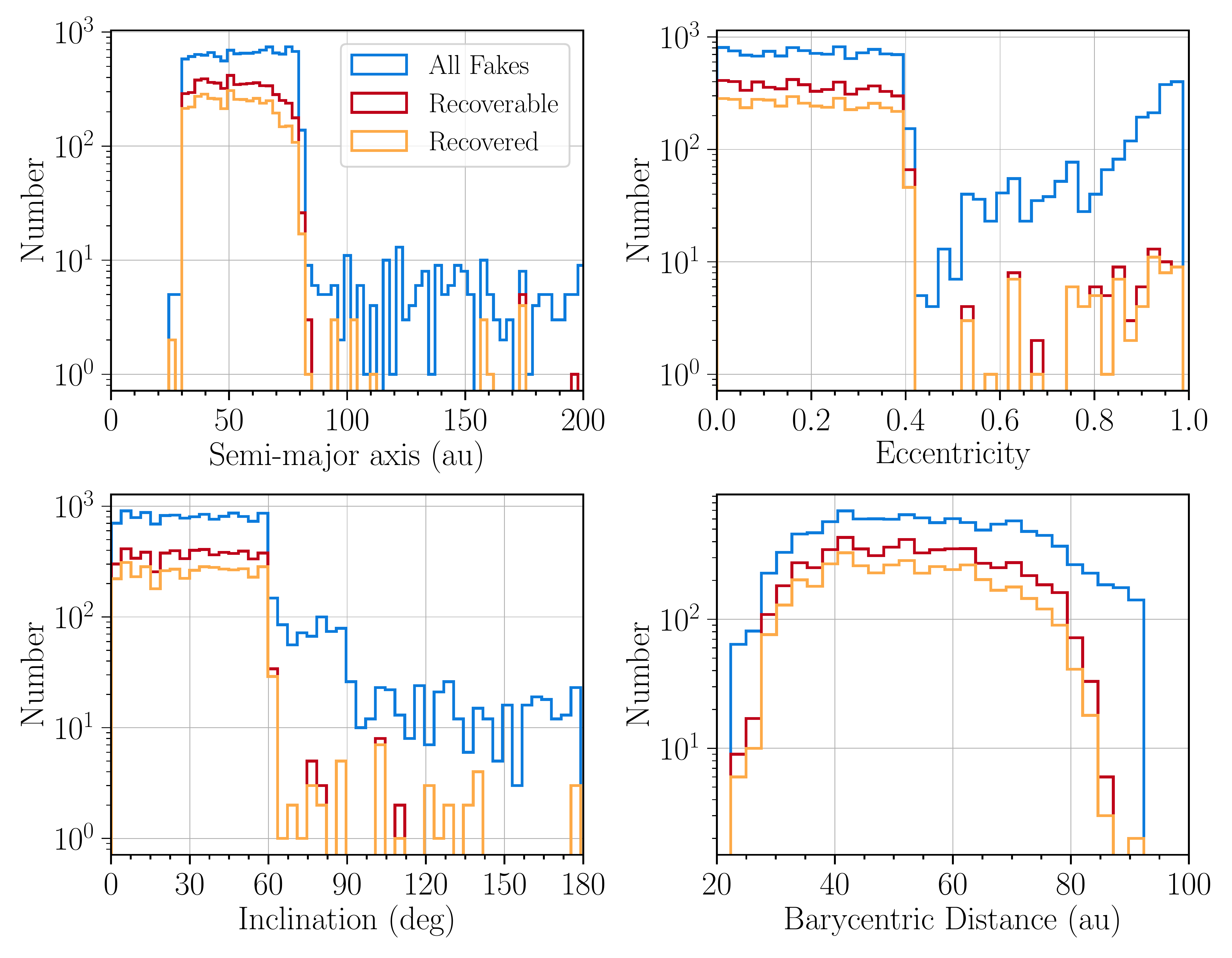}
	\caption{Histograms of semi-major axis $a$, eccentricity $e$, inclination $i$, and barycentric distance $d$ of the injected fakes. The blue lines represent all injected fakes, the red lines (``recoverable'' fakes) represents those with rates $150 < r < 400$ px/day and $m_{VR} < 25.5$, well inside the completeness limit of both rate and magnitude selection functions (as discussed in the Section \ref{sec:joint}). The orange lines indicates those fakes that were recovered by the digital tracking procedure. The histograms of $a$ and $d$ were reduced the region of parameter space with the majority of the objects.\label{im:fakes}}
\end{figure}

{}
The DEEP images were processed using the GPU-based \texttt{KBMOD} algorithm \citep{Whidden2019,Smotherman2021}. A detailed presentation of the full object recovery methodology is given in \cite{DEEPVI} \citep[see also][]{DEEPV}. The detection ``tracklets'' (a pair of sky coordinates derived from the digital tracking procedure projected to the beginning and ending of exposures of each long stare) were subsequently linked into orbits. \change{In \cite{DEEPVI}, we demonstrate that this recovery procedure procedure leads to high confidence recoveries in both injected fakes and known objects, leading to high quality orbit fits ($\sigma_a/a \approx 10^{-4}$) and neglegible contamination.} The digital tracking procedure is conducted on a two dimensional grid of rates of motion $r$ and a positional angle $\phi$, where positive $\phi$ indicates decreasing ecliptic longitude. In the results presented here, our search was limited to $90 < r < 400$ px/day (1 DECam pixel is $0.263\arcsec$) and $|\phi| < 45\deg$. As seen in Figure \ref{im:fakes}, this choice exhausts the parameter space of bound orbits from $20$ to $80$ au, including the entirety of the Kuiper belt. We note that objects at \change{slightly} higher or lower rates \change{can still be recovered} by the search, as the joint fit procedure of \cite{Smotherman2021} corrects for \change{rates approximately outside the search rate that still make it above the detection threshold if it produces a meaningful (but potentially trailed) stack}.

We use the LSST Science Pipelines \citep{Juric2017} in order to inject these fakes into the images. Objects were injected at the sky coordinates derived from their orbits at the midpoint of each exposure, and used the realistic PSF model derived for each exposure \citep{DEEPVI}. These fakes were injected after astrometric and photometric calibration of the images and before the creation of coadded templates and differenced images, guaranteeing that our objects have meaningful PSF fluxes that correspond to their nominal magnitudes at that exposure. This also means that any image processing artifact stemming from the image coaddition and differencing procedures will impact real and injected objects equally. In other words, we have injected a realistic set of synthetic objects into the data that will be detected by our detection pipeline along with the real objects, thus, allowing us to determine the detection efficiency for our survey.

 Our goal is to fully understand the limits of what can be detected in single night stacks and subsequently linked to other nights  in the DEEP survey, so 
 this population is not constrained by the expected structure of the trans-Neptunian region, but rather to fully characterize the range of parameters where we can expect to discover objects. To do this, it is important to understand the efficiencies in the regions of the parameter space where objects have been found and also where no objects have been detected, so this population (intentionally) goes beyond the boundaries of the known TNOs. \change{By constructing such a population model, instead of, for example, using a full model of the trans-Neptunian region, we ensure that we are not biasing ourselves towards what other projects have seen, which would impact our ability to make statistical statements about our observations.} This population is divided in three parts:
\begin{enumerate}
	\item A low eccentricity ($e < 0.3$) population in the $30 < a < 80$ au range. This accounts for the classical Kuiper belt, as well as moderate $a$ and high-$q$ objects. $a$ and $e$ are uniformly distributed in this range, and the inclination distribution is $\propto \sin(i)$ in the $0\degr \leq i \leq 60\degr$ range, leading to an approximately uniform distribution in $i$ for objects inside the survey footprint. The longitude of ascending node $\Omega$, argument of perihelion $\omega$ and mean anomaly $\mathcal{M}$ are uniformly distributed in the $[0\degr, 360\degr]$ range. 
	\item An ``excited'' population with $a$ logarithmically distributed in the $20 < a < 2000$ au range, and perihelia uniformly distributed between $10$ and $100$ au. To avoid creating unphysical orbits, objects where we sampled $a<q$ are changed to $e=0$. The inclinations are distributed as in the first population, but going up to $90\degr$, and $\Omega$ and $\omega$ are uniformly distributed. Objects with $a > 150$ au and $q > 30$ au are limited to $|\mathcal{M}| \leq 30\degr$, to ensure that these objects are not generated at large heliocentric distances. \change{This choice of parameters leads to a population that reproduces the range of parameters for the scattering and detached TNO populations.}
	\item A fully isotropic population that exhausts the parameter space of all bound orbits following the prescription of \cite{Bernardinelli2019}. This population covers the distance range between 25 and 1000 au, with half of the synthetic objects uniformly distributed between $25 < d < 80$ au, and the other half logarithmically distributed between $80 < d < 1000$ au. This population includes objects such as putative distant planets \citep[e.g.][]{Batygin2019} as well as comets \change{and other highly eccentric orbits} at Kuiper belt distances \citep[e.g.][]{Bernardinelli2021a}. 
\end{enumerate}

This population is illustrated in Figure \ref{im:fakes}. We note that each histogram of ``recoverable'' objects (limited to $150 < r < 400$ px/day and $m_{VR} < 25.5$) and ``recovered'' objects qualitatively follow each other closely (up to a normalization), and there is no obvious structure in these distributions, i.e. there is no significant dependency in $e$ and $i$, while the dependency in $a$ and $d$ are directly related to the dependency in $r$ (quantified in Section \ref{sec:eff}). This indicates that our selection function is not dominated by the orbital elements of each object, but rather by the rates and magnitudes. \change{To see that this must be the case, we note that during the four hours of our long stares, the motion of any given TNO is linear and dominated by the reflex motion of the Earth, scaled by the topocentric distance of the object, suppressing the motion driven by the instantaneous barycentric velocity of this object.} In Section \ref{sec:eff} we discuss how these parameters can be characterized.

These objects intentionally span a wide range of apparent magnitudes (20\% of the sample uniformly distributed between $20< m_{VR} < 24$ and 80\% between $24 < m_{VR} < 28$), significantly beyond the detection limits of the survey, so we can have a complete understanding of the efficiencies and potential pitfalls in the analyses. These magnitudes are assigned to the object for Jan 1 2020, and their magnitudes reproduce the expectation given their motion and the motion of the Earth. Half of the objects are also assigned a sinusoidal light curve that changes their magnitude by $\delta m = A \sin( 2 \pi t/T + \phi_0)$, where the \change{semi-}amplitude $A$ is distributed between 0 and 0.5 magnitudes, peaking at $A = 0.25$ mag. The periods $T$ are uniformly distributed between 2 and 100 hours\change{, with a random phases at the Jan 1 2020 epoch; note that for any period significantly longer than the four hours of the long stare, this means that we are sampling the object at a fully randomized phase for each long stare. This simplified model enables a reasonable characterization of the effect of the light curves in the discoverability of our synthetic objects, and reproduces the range seen by \citep{DEEPIV} of periods and amplitudes.} That is, this choice of magnitude amplitudes broadens our efficiency functions in a realistic manner (so that fainter objects are found if they are at the peak of their light curves, and, conversely, brighter objects are missed if their light curve is at a minimum). All reported magnitudes correspond to the mean magnitude of each object during the corresponding long stare. We will defer the full characterization of our light curve amplitude selection function to a future publication (see also \citealt{DEEPIV}).

\section{Estimating the detection efficiency}
\label{sec:eff}

The detectability of an object combines both the information on magnitude $m$ and orbital elements $\mathbf{a} = \{a,e,i,\Omega,\omega,\mathcal{M}\}$ by factoring in the detection probabilities $p(m)$ with the survey geometry and digital tracking efficiencies $p(\mathbf{a}, m|\text{survey geometry, digital tracking})$, and, for objects linked in multiple nights, the linking probability $p(\mathbf{a}, \{ m_\mu \} | \mathrm{linking})$, accounting for the probability that the object recovered in multiple pointing groups $\mu$ is subsequently linked to an orbit \change{and also that such a linkage represents the proper set of detections of the object}\footnote{\change{Note that, for our purposes, we do not require that \emph{all} detections of a given object in the survey are recovered - we only require the recovery of a subset that specifies the object.}}. While we can easily estimate the magnitude-independent $p(\mathbf{a}| \text{survey geometry})$, the digital tracking probability of detection is both a function of magnitude $m$ and rate of motion $r$. This is distinct from a survey that detects objects in each image, as TNOs are typically very close to point sources, and so their detectability is dominated by the the limiting magnitude of the exposure for an object inside its effective area. We introduce here a methodology that simultaneously accounts for the magnitude and rate completeness of each pointing group.

\subsection{Efficiency as a function of magnitude}
We begin by estimating the detection efficiency as a function of magnitude for each pointing group. We follow the prescription of \cite{Bernardinelli2019}, where we parametrize the efficiency by a probability of detection $p(m|\boldsymbol\theta)$ for an object with magnitude $m$ using a set of variables $\boldsymbol\theta$. After the synthetic objects have been properly recovered and identified, we index the recovered objects by $\alpha$ and non-recovered by $\beta$. We maximize the likelihood
\begin{equation}
    \mathcal{L} = \prod_{\alpha} p(m_\alpha | \boldsymbol\theta) \prod_\beta [1 - p(m_\beta|\boldsymbol\theta)]. \label{eq:like}
\end{equation}
This likelihood allows us to estimate the parameters of our chosen efficiency function without having to bin our injected samples.

There are many possible choices of functional forms for $p(m|\boldsymbol\theta)$ \citep[\emph{e.g.}][]{Gladman1998,Bernstein2004,Jones2006,Bannister2015,Bernardinelli2019}. The simplest possible form is a step function, where the only free parameters are the limiting magnitude $m_0$ and the efficiency for $m < m_0$. Smooth functional forms are also possible, with the logistic function
\begin{equation}
    p(m|m_{50}, c, k) = \frac{c}{1 + \exp(k(m - m_{50}))}\label{eq:m50}
\end{equation}
being one of the simplest possible parametrizations. Here, $m_{50}$ is the magnitude of 50\% completeness, $c \in [0,1]$ the peak efficiency and $k$ the transition sharpness between the ``detected'' and ``not detected'' regimes. These parameters are such that $p(m) = c$ for $m \ll m_{50}$, $p(m_{50}) = c/2$ and $p(m) = 0$ for $m \gg m_{50}$. The limit $k \to \infty$ corresponds to a step function. A similar parametrization finds the magnitude of 25\% efficiency ($m_{25}$) fitting two logistics\footnote{Note that, in the literature, this function is sometimes expressed as two hyperbolic tangent functions. This is equivalent to what we show here.} with transition sharpness $k_1$ and $k_2$,
\begin{equation}
    p(m|m_{25}, c, k_1, k_2) = \frac{c}{[1 + \exp(k_1(m - m_{25}))] [1 + \exp(k_2(m - m_{25}))]}.
    \label{eq:m25}
\end{equation}
We also posit a third model, that we fit as a sanity check to verify whether we are overfitting our data, that uses three logistics and has a characteristic magnitude of $12.5\%$ completeness $m_{12.5}$. We note that this methodology extends to other functional forms, not just the ones presented here \citep[see, e.g.,][for a functional form that quantifies lost bright objects]{Bannister2015}.

To decide the optimal functional form, we use the Bayesian information criterion \citep[BIC,][]{Schwarz1978} test on the maximum likelihood $\mathcal{L}$. For $N_\mathrm{par}$  model parameters and $N_\mathrm{data}$ data points, we have that the optimal model minimizes
\begin{equation}
    \mathrm{BIC} = N_{\mathrm{par}} \log N_\mathrm{data} -2\log \mathcal{L} .
\end{equation}
Since we are interested in the general behavior of our efficiency function, we perform this by fitting one $p(m)$ to the entire set of synthetic objects, that is, we estimate an approximate selection function for all long stares. We restrict our data to $150 < r < 400$ px/day, a region where the dominating cause of object loss is from the magnitude selection function. Of the three functional forms for $p(m)$ presented here, Equation \ref{eq:m25} has the lowest BIC of $\mathrm{BIC}=8055.14$, while Equation \ref{eq:m50} and the triple exponential have $\mathrm{BIC}=8071.99$ and $\mathrm{BIC}=8061.12$ respectively. Following \cite{Kass1995}, the difference between two BIC values approximates the Bayes factor between these two models, and therefore, among a set of models (model 1, model 2, and model 3), we can estimate the odds that a given model best fits our data. If we call model $i$ the model with the lowest BIC, then the odds $\mathcal{O}$ that model that best fits the data is approximately
\begin{equation}
    \mathcal{O}_{i, j\neq i} = \exp \left[ (\mathrm{BIC}_i-\mathrm{BIC}_j)/2\right].
\end{equation}
The odds against Equation \ref{eq:m50} fitting the data better than Equation \ref{eq:m25} are approximately 4570:1 and the odds against a triple exponential fitting the data better than Equation \ref{eq:m25} are approximately 20:1. We have therefore confirmed that the model given by Equation \ref{eq:m25} is likely the best model for our application of the three considered here.

However, we stress that the BIC is an approximate metric and that the choice of model among these three models does not qualitatively alter our characterization, as the $m_{50}$ and $m_{25}$ models predict similar drop-off points. Converting the best-fit $m_{25} = 26.22$ from Equation \ref{eq:m25} to a $m_{50}$ value, we get a converted $m_{50}$ of 25.96, while fitting Equation \ref{eq:m50} directly leads us to $m_{50} = 25.93$. Our results are therefore only weakly sensitive to the choice of $p(m)$. From now on, we will assume $p(m)$ is given by Equation \ref{eq:m25}, but our results and methodology are independent of this choice.

\subsection{Joint rate and magnitude selection function}
\label{sec:joint}
The selection function for each pointing group is not only a function of the source magnitude $m$, but also the rate of motion $r$, which we assume has the same functional form and parameters for all pointing groups. Thus, any fit of Equations \ref{eq:m50} and \ref{eq:m25} do not yield a meaningful completeness $c$, as, in principle, one can add an arbitrary number of bright objects outside the rate range in which objects are meaningfully detected, or reduce the samples to a predefined range of rates. On the other hand, attempting to fit a \emph{rate} completeness function requires the data to be limited to $m \ll m_{25}$, also necessarily reducing the sample size. We propose a methodology that accounts for both rate and magnitude selection functions simultaneously. We posit a simple smooth functional form for the rate completeness, 
\begin{equation}
	p(r|r_{50,1}, r_{50,2}, \kappa_1, \kappa_2, r_0) = 
	\begin{cases} \frac{1}{1 + \exp(\kappa_1 (r - r_{50,1}))}, & r < r_0
	\\ \frac{1}{1 + \exp(\kappa_2 ( r - r_{50,2}) )}, & r > r_0 
	\end{cases} ,
\end{equation}
corresponding to a logistic function in each end of the distribution (with $\kappa_1 < 0$ and $\kappa_2 > 0$), and a rate of 50\% completeness for each side. The constant $r_0$ determines the transition between the two regimes, and can be chosen prior to the fit without loss of constraining power. \change{In principle, we could attempt a similar statistical exercise as in the previous section to determine the optimal functional form for this function. In practice, the recovery efficiency drops to zero quite steeply, so we can safely posit this simpler form.} We assume that this rate selection function is the same for all pointing groups, as the underlying digital tracking algorithm in \texttt{kbmod} does not change in each processing. 

\begin{figure}[ht!]
	\centering
	\includegraphics[width=0.495\textwidth]{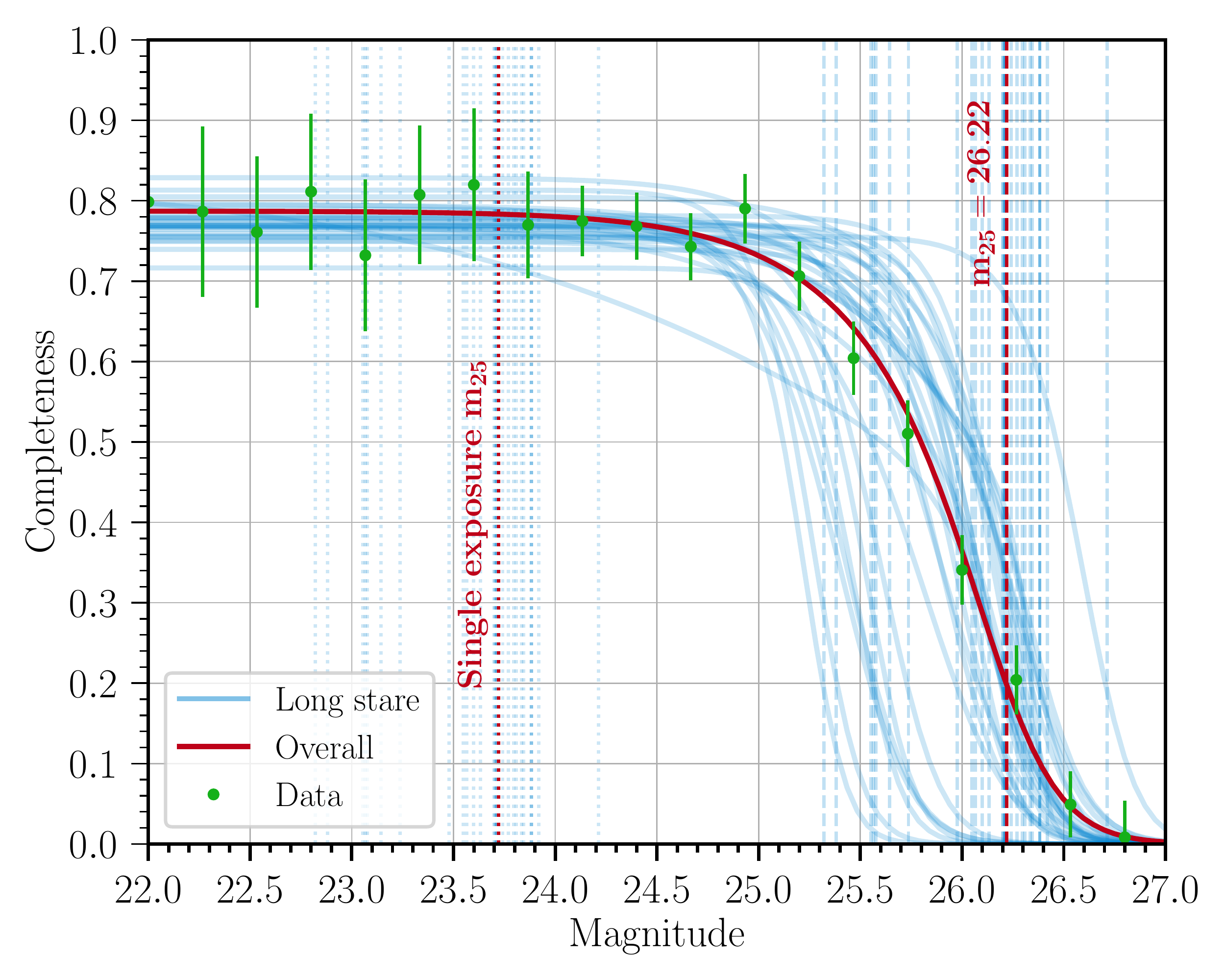}
	\includegraphics[width=0.495\textwidth]{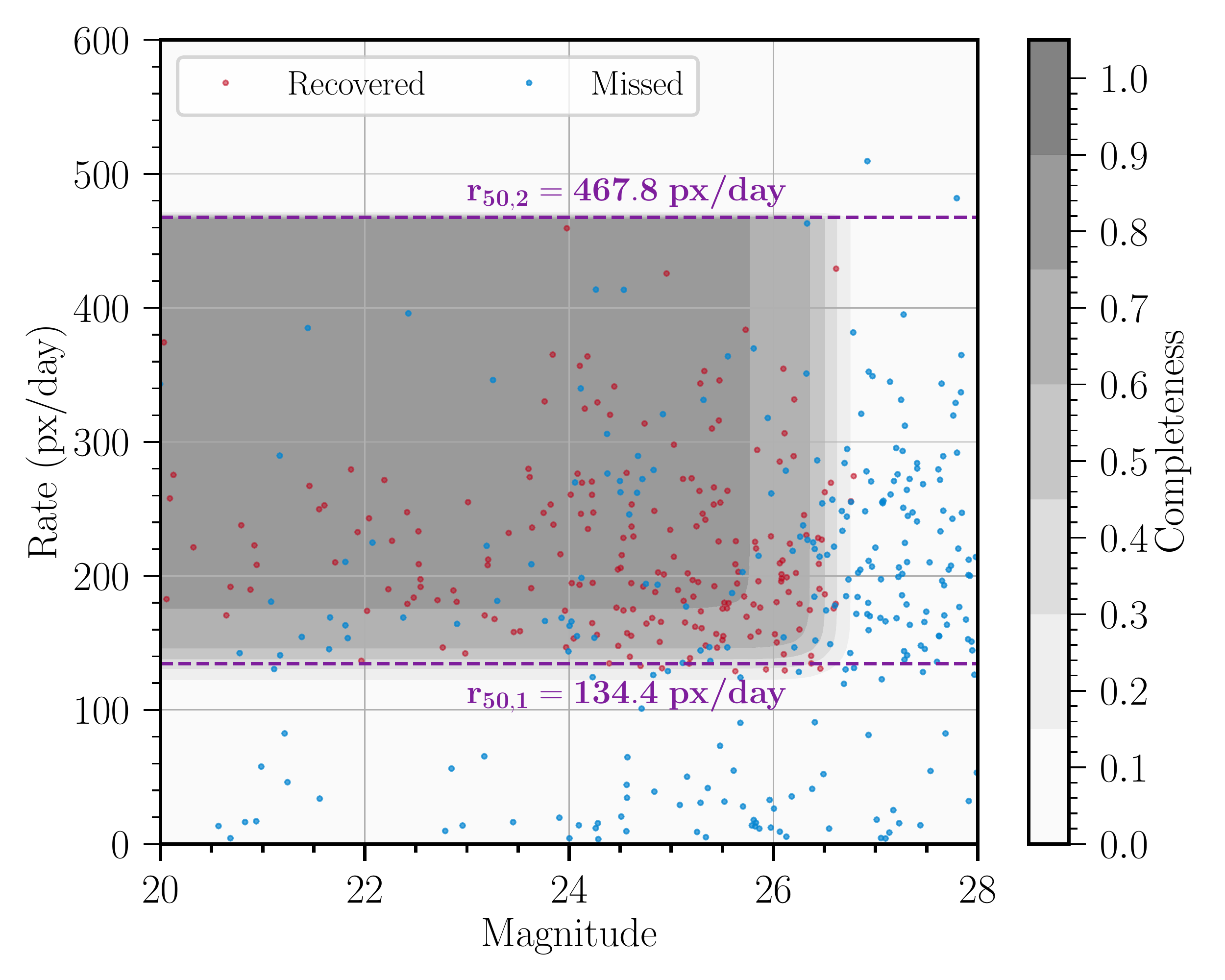}
	\caption{\emph{Left:} Completeness versus magnitude for all long stares (solid blue) as well as the overall completeness for the survey (solid red), where all pointing groups are fit to the same probability curve. The vertical dashed lines indicate the best-fit $m_{25}$ for each long stare (blue) and for the entire data (red). The dotted lines represent the $m_{25}$ each exposure in the long stare (blue) and the survey (red) would have if the data were not stacked. The green dots are the true binned recovery rates averaged over all long stares with objects limited to $r_{50,1} < r < r_{50,2}$ with Poisson error bars. \emph{Right:} Rate and magnitude for all synthetic sources (recovered in red and missed in blue) in our long stare with the deepest $m_{25}$. The shaded grey curves represent the completeness in Equation \ref{eq:joint}, and the horizontal dashed purple lines represent the 50\% rate completeness terms ($r_{50,1}$ and $r_{50,2}$).\label{im:completeness}}
\end{figure}

The combined rate and magnitude selection function becomes
\begin{equation}
	p(m,r | m_{25}, k_1, k_2, r_{50,1}, r_{50,2}, r_0, \kappa_1, \kappa_2, c) = c \cdot p(m|m_{25}, k_1, k_2, 1) \cdot p(r|r_{50,1}, r_{50,2}, \kappa_1, \kappa_2, r_0).
	\label{eq:joint}
\end{equation}
As before, $c$ corresponds to the peak efficiency (when $m \ll m_{25}$ and $r_{50,1} < r < r_{50,2}$), the parameter range where objects are systematically recovered. To simplify the notation, we index each pointing group by $\mu$ and define $\boldsymbol\theta_\mu \equiv \{m_{25,\mu}, c_\mu, k_{1,\mu} k_{2,\mu}\}$, $\boldsymbol\varphi \equiv \{ r_{50,1}, r_{50,2} \kappa_1, \kappa_2\}$ (note that $\boldsymbol\varphi$ is independent of pointing group). We generalize the likelihood in Equation \ref{eq:like} by including the rate in addition to the magnitude of each implanted source and by multiplying it over all pointing groups, so
\begin{equation}
	\mathcal{L} = \prod_\mu \mathcal{L}_\mu  = \prod_\mu \left[ \prod_{\alpha} p(m_\alpha, r_\alpha | \boldsymbol\theta_\mu, \boldsymbol\varphi) \prod_\beta [1 - p(m_\beta, r_\beta|\boldsymbol\theta_\mu, \boldsymbol\varphi)] \right]. \label{eq:ml}
\end{equation}
Maximizing this $\mathcal{L}$ achieves our goal of obtaining a meaningful $\{c_\mu\}$ as well as characterizing the global rate selection function. \change{One way of understanding Equation \ref{eq:ml}, then, is that for any given long stare, the rate selection function is informed by all other long stares, and so its constraints are stronger than what could be derived from any individual long stare, mitigating the effect of the joint dependency of magnitude selection function with the rates of the implanted sources. That is, we do not need as many sources in each $(m,r)$ bin as we would if we were trying to reconstruct the joint rate and magnitude selection functions for each individual long stare.} We illustrate this procedure in Figure \ref{im:completeness}. 

\section{Simulation methodology}
\label{sec:methods}
The characterization in the previous sections established that, for any given TNO inside our survey area, our discoverability is a function of its on-sky motion (the rate) and magnitude. Thus, these are the primary parameters that our survey simulator needs to decide which objects are recovered. For any given object, the simulation proceeds as follows:
\begin{enumerate}
	\item The object's orbital elements or state vector are integrated to all exposure times for the survey, and on-sky positions are generated. These positions are then used to check whether an object could be seen by a functional DECam CCD with this pointing. This properly accounts for chip gaps, and so this information is also used to decide whether this synthetic object would lie inside its search CCD for long enough to be properly recovered by the digital tracking methodology. Note that only the synthetic objects that have successfully passed this check are included in the procedure of Section \ref{sec:eff}.
	\item For all individual observations that could be recovered by the digital tracking procedure in a given long stare $\mu$, we compute its rate $r$ and mean magnitude $m$ given the parent object's light curve (that can be constant), and sample a random uniform number $u \in [0,1]$. If $p(m,r|\boldsymbol\theta_\mu,\boldsymbol\phi) > u$, the object is considered to be recovered. Since $0 \leq p(m,r|\boldsymbol\theta_\mu,\boldsymbol\phi)  \leq c_\mu$, even bright objects have a $\sim 20 \%$ chance of being discarded (as $\langle c_\mu \rangle \approx 0.8$), thus properly accounting for the area lost in the digital tracking procedure. 

	\item We compute the discoverability of the orbit given its arc (the time span between the first and last observations); the shortest arc after dropping either the first or the last night of data; and the number of nights in which the object was detected. These inform whether the object would be linked or not. For the processing of \cite{DEEPVI}, an object is linked with high efficiency ($94\%$) if it was observed in at least four nights, with complete arcs spanning 0.8 years (0.5 years for the shortened arcs). We sample another uniform number $u \in [0,1]$, and compare to the linking efficiency $\ell \approx 94\%$, to determine whether the object would be properly linked. This accounts for objects with the proper number of observations but whose detections are astrometric outliers (that is, the recovered positions do not agree with the nominal position for the object) or objects whose orbit fits \change{yield poor results, as measured by their reduced $\chi^2$ (see \citealt{DEEPVI} for a detailed discussion).}

\end{enumerate}

This \change{simulation is implemented as an extension of} \texttt{DESTNOSIM} \citep{Bernardinelli2022}, the survey simulator initially developed for the Dark Energy Survey (DES) search for TNOs \citep{Bernardinelli2019,Bernardinelli2022}, and makes use of the dynamics package \texttt{Spacerocks} \citep{Napier_spacerock} \change{to produce the ephemerides for the synthetic objects}. The modular nature of the software allows characterization of any processing of DEEP, as well as any similar survey, such as DES. A full tutorial, as well as documentation and scientific applications of the software, are publicly available on a \texttt{GitHub}\footnote{\url{https://github.com/bernardinelli/DESTNOSIM}} repository. We encourage users to extend the capabilities of the software, as well as publicly release their analyses in addition to their publications, ensuring that their results are reproducible by the community. 

\section{Applications}
\label{sec:numbers}
As an initial demonstration of the software, we study two simple scenarios with toy populations that quantitatively illustrate its simulation capabilities. 
\subsection{Effective search area}

\begin{figure}[h]
	\centering
	\includegraphics[width=0.495\textwidth]{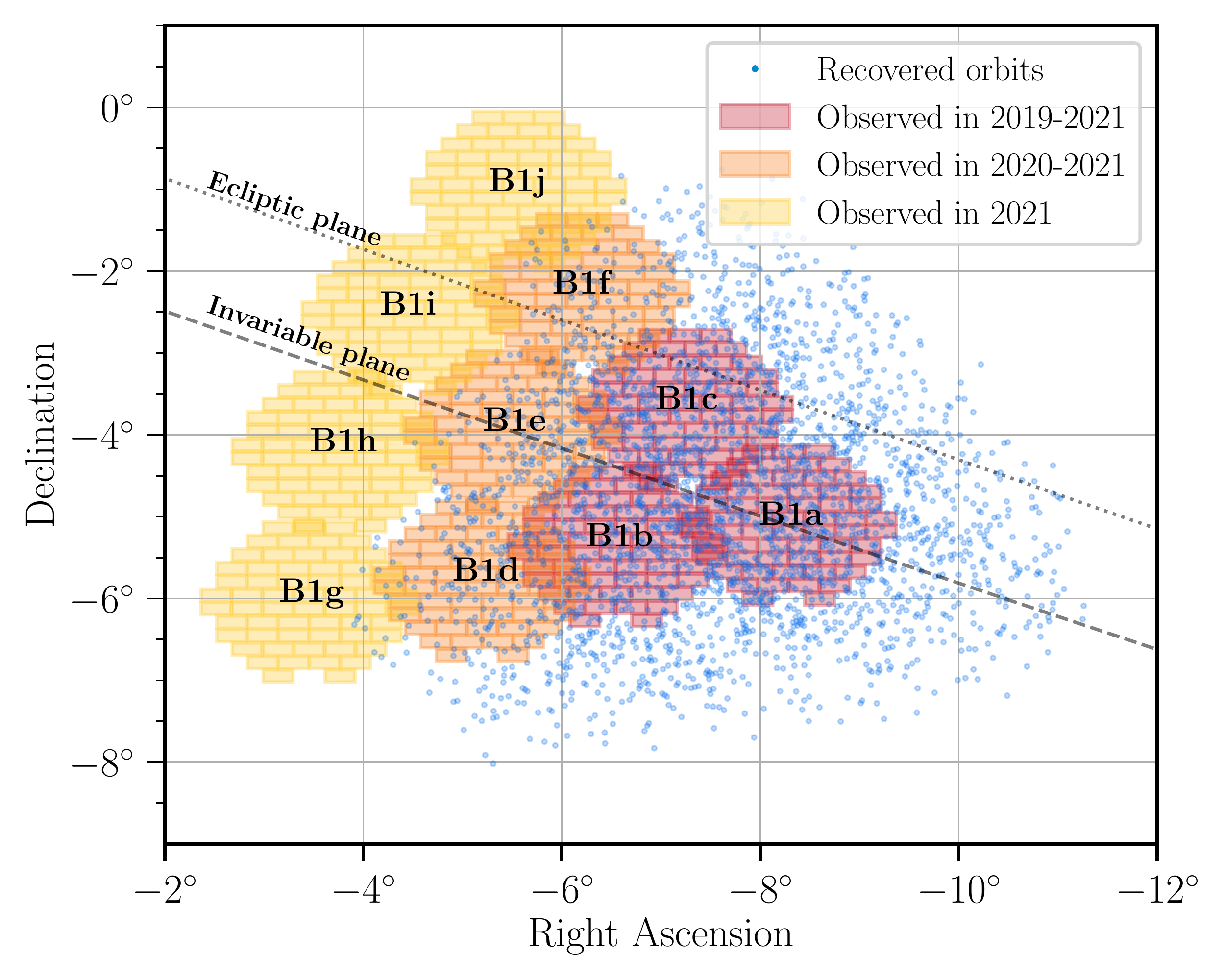}
	\caption{Sky coordinates of the B1 fields. The shaded regions indicate the observing coverage of each pointing, with the colors representing the observing years that covered each pointing (red: 2019-2021, orange: 2020-2021, yellow: 2021). The blue dots show a representative sample (20\%) of the TNO that could be successfully linked and are displayed in their nominal sky position in 2020-01-01 (the reference epoch). The dotted (dashed) line mark the location of the ecliptic (invariable) plane. We note that few objects in fields B1g-B1j are recovered: this is expected, as these fields were not observed in 2019 and 2020, so objects from these fields have not yet been imaged to be linkable in the data.\label{im:area}}
\end{figure}

The 2019-2021 data from the DEEP B1 field include 10 distinct DECam pointings (see Figure \ref{im:area}). Using the camera's nominal field-of-view ($3\deg^2$), it would seem that our search area for multi-year orbits is $\approx30\deg^2$. This is, however, not precise: not all fields were observed in all years, not all of DECam's CCDs are functional, and there is a fraction of area lost due to the survey's observing geometry (the combination of fields, cadence and camera footprint). The observing strategy of DEEP is such that fields are imaged in an expanding ``cone'' pattern: in 2019, the B1a, B1b and B1c fields were visited; in the 2020, the B1a through B1f fields were observed, and all fields were observed in 2021. 

We use the simulator to estimate the ``effective'' search area, that is, the area in the sky that any potential bound orbit could be observed by DEEP and then subsequently be recovered by the linking procedure. For these purposes, we ignore the magnitude and rate selection functions, as our interest is in the geometrical design of the survey. We simulate an isotropic, all-sky population of objects as defined above, but we fix their an initial distance to 40 au (note that, as defined here, this effective area is distance dependent, as we are accounting for the motion of the object). The procedure of \cite{Bernardinelli2019} distributes orbits in a spherical Fibonacci lattice, and so the fraction of the points we recover is a proxy for the area on the sphere that is available for objects to be searched (see \citealt{Gonzalez2009} for a detailed explanation of these lattices). We simulated $4\cdot10^7$ TNOs and, of these, 14408 satisfy the linking criteria of Section \ref{sec:methods}, leading to $A_{\mathrm{eff}} \approx 14.8\deg^2$. We emphasize that this effective area will change as the survey progresses.

\subsection{Limits on a distant, dynamically cold population}
There are few objects known in the outer Kuiper belt, defined as $a$ larger than Neptune's 2:1 mean motion resonance and low eccentricities \citep[$e \lesssim 0.24$,][]{Gladman2021}. While the high inclination component of this population extends as far as the 3:1 resonance \citep{sheppard2016,Bernardinelli2022}, this component is thought to be created by a combination of mean motion resonances and the Kozai effect \citep{Gomes2008}, and no dynamically cold ($e \leq 0.24$ and $i \leq 5\degr$) objects with $a>60$ au are known. 
\change{In \cite{DEEPVI}, we recovered 110 TNOs, primarily cold Classicals \citep[low inclination, low eccentricity objects between Neptune's 3:2 and 2:1 mean motion resonances, see Figure 7 of ][]{DEEPVI}. In agreement with other surveys, we also also did not detect any objects in this high-$a$, low-$e$ and low-$i$ region of the parameter space, so we can} use our simulator to place limits on the existence of a distant thin disk of objects past this $a$ range. We emphasize that this is not the first time such limits have been placed \citep{Trujillo2001,Trujillo2001a,Allen2002,Bernstein2004}, but no limits with our combination of area and magnitudes have been placed to date, and such a population has not been entirely ruled out \citep[see Section 6.2.3 of][]{Gladman2021}. 

We simulate a population similar to the stirred classical Kuiper belt population of \cite{Petit2011}, \change{following their $a \propto a^{-5/2}$ surface density for the Kuiper belt, but moving it further to the $60 \leq a \leq 80$ au range}. The other parameters follow \cite{Petit2011}: $e$ uniformly distributed between $0$ and $0.03$; $i \propto \sin(i) \exp(-i^2/(2\sigma_i^2))$, where $\sigma_i = 2.6\degr$ and $\omega,\Omega,\mathcal{M}$ uniform in the $0\degr$ to $360\degr$ range. 
We simulate this population in the $24 \leq m \leq 26$ range, \change{corresponding roughly to $5 < H < 8$ at these distances}. Note that, for $m \ll 24$, our magnitude efficiency is constant, and our population estimates are limited by our survey area, so we do not need to simulate brighter objects. We measure the fraction of members of this population recovered at each magnitude bin $m_b$. This allows us to compute our probability of detecting an object $p(m_b)$. Assuming a size distribution, we can convolve it with our detection probability $p(m_b)$ to place limits on the number of objects with an absolute magnitude $H \leq H_b$. Using Poisson statistics, we have that an undetected population whose simulation leads to 3 detections on average can be rejected it with $95\%$ confidence, as the Poisson error bar is $\sqrt{3}$ for such a process. Assuming the \cite{Fraser2014} size distribution for the cold Classical population, \emph{namely, a broken power law with slope $1.5$ for the larger objects and $0.3$ for the smaller objects, with the break at $H = 6.9$}, we have $N(H \leq 8) = 8 \times 10^3$ objects in this region of the outer Solar System. It's important to note that any size distribution that is \emph{shallower} than that of \cite{Fraser2014} would lead to a stronger constraint, that is, a smaller number of objects. A more useful limit would be for $a \sim 90$ au \citep{Gladman2021}, but our completeness drops sharply after $d \approx 80$ au (see Figure \ref{im:fakes}). Further data releases of the DEEP survey will attempt to recover and place limits to these more distant populations.

\section{Summary}
\label{sec:conc}
We constructed a survey simulator for the DEEP survey via injection of a population of synthetic outer Solar System objects that exhausts the parameter space of orbits at Kuiper belt distances (and beyond) into the DEEP images. This software, publicly released on \texttt{GitHub}, enables the user to characterize our observational biases as a function of magnitude and rate of motion for each pointing group, and of the survey as a whole. Such a construction enables further studies of recoverability of populations of trans-Neptunian objects by the survey, and thus detailed comparisons between models of the structure of the trans-Neptunian region and the TNOs observed by the DEEP survey. In addition to the software release, we will provide real world use cases in the form of Jupyter Notebooks, ensuring that our analyses are reproductible by the community.

In this first release, we focused on the 2019-2021 data from the B1 fields \citep{DEEPVI} processed through \texttt{kbmod} \citep{Whidden2019,Smotherman2021}, but the methodology developed here is general and can be extended to any survey that uses digital tracking to recover TNOs. Our lack of dynamically cold objects in distant ($a>60$ au) orbits combined with the survey simuator enables us to derive that this yet undetected potential population has at most $8 \times 10^3$ members with $H \leq 8$.

\bibliography{references}
\bibliographystyle{aasjournal}

\begin{acknowledgements}
This work is based in part on observations at Cerro Tololo Inter-American Observatory at NSF’s NOIRLab (NOIRLab Prop. ID 2019A-0337; PI: D. Trilling), which is managed by the Association of Universities for Research in Astronomy (AURA) under a cooperative agreement with the National Science Foundation.

This work is supported by the National Aeronautics and Space Administration under grant No.\ NNX17AF21G issued through the SSO Planetary Astronomy Program and by the National Science Foundation under grants No.\ AST-2009096 and AST-2107800. This research was supported in part through computational resources and services provided by Advanced Research Computing at the University of Michigan, Ann Arbor. This work used the Extreme Science and Engineering Discovery Environment \citep[XSEDE; ][]{XSEDE}, which is supported by National Science Foundation grant number ACI-1548562. This work used the XSEDE Bridges GPU and Bridges-2 GPU-AI at the  Pittsburgh Supercomputing Center through allocation TG-AST200009.

H. Smotherman acknowledges support by NASA under grant No.\ 80NSSC21K1528 (FINESST). H. Smotherman, M. Juri\'{c} and P. Bernardinelli acknowledge the support from the University of Washington College of Arts and Sciences, Department of Astronomy, and the DiRAC Institute. The DiRAC Institute is supported through generous gifts from the Charles and Lisa Simonyi Fund for Arts and Sciences and the Washington Research Foundation. M. Juri\'{c} wishes to acknowledge the support of the Washington Research Foundation Data Science Term Chair fund, and the University of Washington Provost’s Initiative in Data-Intensive Discovery. 

This project used data obtained with the Dark Energy Camera (DECam), which was constructed by the Dark Energy Survey (DES) collaboration. Funding for the DES Projects has been provided by the US Department of Energy, the US National Science Foundation, the Ministry of Science and Education of Spain, the Science and Technology Facilities Council of the United Kingdom, the Higher Education Funding Council for England, the National Center for Supercomputing Applications at the University of Illinois at Urbana-Champaign, the Kavli Institute for Cosmological Physics at the University of Chicago, Center for Cosmology and Astro-Particle Physics at the Ohio State University, the Mitchell Institute for Fundamental Physics and Astronomy at Texas A\&M University, Financiadora de Estudos e Projetos, Funda\c{c}\~{a}o Carlos Chagas Filho de Amparo \'{a} Pesquisa do Estado do Rio de Janeiro, Conselho Nacional de Desenvolvimento Cient\'{i}fico e Tecnol\'{o}gico and the Minist\'{e}rio da Ci\^{e}ncia, Tecnologia e Inova\c{c}\~{a}o, the Deutsche Forschungsgemeinschaft and the Collaborating Institutions in the Dark Energy Survey.

The Collaborating Institutions are Argonne National Laboratory, the University of California at Santa Cruz, the University of Cambridge, Centro de Investigaciones En\'{e}rgeticas, Medioambientales y Tecnol\'{o}gicas–Madrid, the University of Chicago, University College London, the DES-Brazil Consortium, the University of Edinburgh, the Eidgen\"{o}ssische Technische Hochschule (ETH) Z\"{u}rich, Fermi National Accelerator Laboratory, the University of Illinois at Urbana-Champaign, the Institut de Ci\`{e}ncies de l’Espai (IEEC/CSIC), the Institut de Física d’Altes Energies, Lawrence Berkeley National Laboratory, the Ludwig-Maximilians Universit\"{a}t M\"{u}nchen and the associated Excellence Cluster Universe, the University of Michigan, NSF’s NOIRLab, the University of Nottingham, the Ohio State University, the OzDES Membership Consortium, the University of Pennsylvania, the University of Portsmouth, SLAC National Accelerator Laboratory, Stanford University, the University of Sussex, and Texas A\&M University.

\end{acknowledgements}
\end{document}